\documentclass[aps,prb,amsmath,amssymb,reprint,superscriptaddress,floatfix,showpacs]{revtex4-2}
\usepackage[english]{babel}
\usepackage[utf8x]{inputenc}
\usepackage[T1]{fontenc}
\usepackage{bm}

\usepackage{xcolor}
\usepackage[left=2.cm,right=2.cm,
    top=3cm,bottom=3cm,bindingoffset=0cm]{geometry}

\usepackage{graphicx}
\usepackage{notes2bib}
\usepackage{amsfonts}
\usepackage{amssymb}
\usepackage{float}
\usepackage{amsmath}
\usepackage{subcaption}
\usepackage{hyperref}

\usepackage[disable]{todonotes}

\begin{document}

\title{Lifetime, collapse and escape paths for hopfions in bulk magnets with competing exchange interactions}

\author{I. S. Lobanov}
\affiliation{Faculty of Physics, ITMO University, 197101 St. Petersburg, Russia}

\author{V. M. Uzdin}
\affiliation{Faculty of Physics, ITMO University, 197101 St. Petersburg, Russia}

\begin{abstract}
The lifetimes of magnetic hopfions on a discrete lattice with competing exchange interactions are calculated
within the framework of the transition state theory for magnetic degrees of freedom.
Three sets of discrete model parameters corresponding to the same continuous micromagnetic model are considered.  
Minimal energy paths for hopfion collapses were found on the  multidimensional energy surface of the system. 
The activation energies of the collapse processes have been calculated. 
It turned out that the activation energy differs significantly for the three considered values of the parameters,
which indicates the importance of lattice effects, when the hopfion radius equals several lattice constants. 
Along with the collapse, the hopfion escape process through the sample boundary is studied. 
It is shown that this process does not require an activation energy. 
The lifetimes of hopfions are found and it is shown that they can exist only at temperatures 
of a few kelvins and practically cannot be generated due to thermal fluctuations.
\end{abstract}

\maketitle


\section{Introduction}
\label{sec:introduction}
Recently, much attention has been attracted by topological structures of small spatial size, the properties of which do not change under continuous transformations of the corresponding states or external perturbations affecting these states.
In two-dimensional magnets, such topological systems include skyrmions, which are considered as promising candidates for elements of a new generation of  racetrack magnetic memory  and neuromorphic devices \cite{Wiesendanger16,Fert17,Song20}.
For these systems, one can introduce an integer topological charge, the geometric meaning of which is the number of times that the unit vectors along the magnetization at each point, built from one center, cover the unit sphere. The invariance of the topological charge with a continuous change in the magnetization implies topological stabilization of such systems \cite{Nagaosa13}.
However, for magnetic moments localized at the nodes of a discrete lattice, topological considerations are strictly speaking inapplicable, and topological stabilization should manifest itself in the values of the activation barrier and the preexponential factor (PF) in the Arrhenius law for the lifetime, which can be obtained within the harmonic approximation of transition state theory (TST) \cite{Bessarab12, Lobanov21}.

Three-dimensional (3D) topological structures are even more diverse and often encountered. Topological defects, such as Bloch points and lines, localized topological solitons, domain walls, skyrmion tubes, can appear both in bulk material and on its surfaces and interfaces \cite{Rybakov15,Ackerman17,Tambovtsev22}. For 3D structures, new types of topological indices can be introduced.
These are also integers, which do not change with continuous deformations of the magnetization or the director in liquid crystals. As in the two-dimensional case for topological solitons on a discrete lattice, the presence of topological indices should manifest itself through the large activation energy of collapse and/or small PF.

 We investigate the structure and stability of the 3D  topological Hopf soliton (hopfion) \cite{Sutcliffe18,Rybakov22,Chen13}. To describe hopfion's topological characteristics, one can define the topological Hopf invariant which is the number of 
 \todo{Is that a standard definition?}
 engagements of rings corresponding to the constant direction of the moments in the magnetic structure. This is an integer that cannot be changed in the bulk material without creating 
 \todo{Other types of defects?}
 Bloch points, resulting in an infinite energy density in the continuous case. 
 Therefore, topological stabilization can also be expected in a 
 sufficiently dense discrete lattice.

Most studies of Hopf solitons are related to the investigation of structures formed in a chiral medium. Chirality can be related to the bulk Dzyaloshinskii-Moriya interaction (DMI) in ferromagnets or DMI induced at the interface with a heavy metal. For such structures, depending on the direction of the DMI vector, N\'eel and Bloch hopfions can be formed 
\cite{Khodzhaev22}. The dynamics of such topological solitons under the action of an electric current has been studied.
 Distinct from 2D ferromagnetic skyrmions, hopfions do not show Hall effects under current. N\'eel-type hopfions move along the current direction via both spin-transfer torques (STTs) and spin-Hall torques (SHTs), while Bloch-type hopfions move either transverse to the current direction via SHT or parallel to the current direction via STT \cite{Wang19}.
The hopfion’s locally uncompensated emergent field leads to a topological Hall signature, although the topological Hall effect vanishes on the global level. 
This however can be used to detect hopfions \cite{Gobel20}.

Micromagnetic modeling shows  that under the influence of  external magnetic fields, hopfions can transform into torons \cite{Li22,Raftrey21}. 
The difference between the eigenmodes of the hopfions  and torons makes it possible to identify and separate these states \cite{Li22,Khodzhaev22,Raftrey21}.

Chiral hopfions have been experimentally observed in magnetic materials \cite{Kent21} and liquid crystal systems \cite{Chen13}.
In a chiral medium, the ground state may not be ferromagnetic, but helical or conical. Localized hopfion states are also observed against the background of such non-collinear states \cite{Voinescu20,Tai22}. Their stability, as in the case of two-dimensional skyrmions, depends significantly on DMI or chiral interactions in liquid crystals.
They can also be stabilized by surface anisotropy and the geometry of nanostructures \cite{Tai-PRL18}. It was shown that such solitons can be electrically and magnetically switched between states with the same or different Hopf indices \cite{Tai18}.

Thus, most theoretical and experimental studies are related to the properties of hopfions in materials with chiral interaction. 
However, they can be obtained even in the absence of these interactions. Micromagnetic modeling shows that such structures can 
appear in toroidal nanoparticles under the action of an Oersted magnetic field. This is due to magnetostatic and exchange 
competition under the action of an azimuthal magnetic field \cite{Castillo21}. 
Hopfions can also be stabilized in frustrated magnets at a certain ratio of exchange interactions between magnetic moments.
They should exhibit rich dynamics, including longitudinal motion along the current direction, transverse motion perpendicular 
to the current direction, rotational motion, and dilation \cite{Liu20}.
The effective field generated by a hopfion, contains magnetic octupole component. This locally uncompensated 
emergent field leads to a new topological Hall peculiarities \cite{Liu22}. 
For the stability of the hopfions and the observation of all these effects, the competition of exchange interactions is sufficient.

Similar stabilization mechanisms exist for skyrmions in frustrated two-dimensional magnets without taking into account of DMI and spin-orbit interaction effects  \cite{Hu17,Heil19}.
In what follows, we will focus on this case. In the continuous model, a frustrated magnet is described by an energy density that contains,
in addition to the square of the magnetization gradient, other contributions with second-order derivatives in space \cite{Rybakov22,Sallermann22}.

On a discrete lattice, such a system corresponds to a model with several exchange parameters depending on the distance between nodes. At the same time, several discrete models with different parameter values can correspond to the same continuous model.
Note that this also applies to the choice of parameters of skyrmion states on a discrete lattice, corresponding to one continuous model or calculations from first principles \cite{Potkina22,Hoffmann20}.

To estimate the stability and lifetimes of 3D hopfions, we will use harmonic TST \cite{Bessarab12,Lobanov21}.
This approach involves the construction of the energy surface of the system as function of parameters that uniquely specify the magnetic state. The Cartesian coordinates of the magnetic moments are used as such parameters, and the condition of constant value of moments within Heisenberg-like theory is taken into account by introducing the Lagrange multipliers \cite{Lobanov21,Lobanov-CPC21}.
The minimum energy path (MEP) between the state corresponding to the hopfion and the homogeneous ferromagnetic state determines the most probable transition scenario and the activation energy of hopfion collapse. The collapse inside the sample and escape through its boundary will be considered.
In contrast to the case of two-dimensional skyrmions \cite{Lobanov16}, 
the MEP that determines the activation energy, passes through a state with Bloch points, 
\cite{Birch21}. PF is calculated  within harmonic TST on the basis of method developed in ref. \cite{Lobanov-CPC21}.

\section{Hopfions in discrete lattice models}
\label{sec:model}
In the semiclassical model of the Heisenberg type, which we follow here, magnetic configuration is described by the vector field of magnetization $\bm M(\bm r)$.
The absolute value of the magnetization for homogeneous medium is
assumed to be constant 
$M_S$, thus it is convenient to introduce
magnetization direction vector $\|\bm n(\bm r)\|=1$: $\bm M(\bm r)= M_s \bm n(\bm r)$, which
totally determines the magnetic state. 

In the discrete model, the exchange interaction is given by a set of exchange integrals depending on the distance between the magnetic moments. Taking into account the interaction beyond the nearest neighbors in two-dimensional systems can significantly increase the estimates lifetimes of magnetic skyrmions \cite{Malottki2017} and even lead to the formation of locally stable skyrmions in the absence of chiral DMI \cite{Hu17,Heil19}.
In micromagnetic models, the exchange interaction is described by terms containing derivatives of magnetization. We will use following advanced functional of micromagnetic energy \cite{Rybakov22}
\[
\omega (\bm r) =
\mathcal{A}\left(\frac{\partial\bm n}{\partial r_\alpha}\right)^2
+\mathcal{B}\left(\frac{\partial^2\bm n}{\partial r_\alpha^2}-\frac{\partial^2\bm n}{\partial r_\beta^2}\right)^2
+\mathcal{C}\left(\frac{\partial^2\bm n}{\partial r_\alpha\partial r_\beta}\right)^2,
\]
Here indices $\alpha$ and $\beta$ run through $x, y, z$, and summation over $\alpha$ and $\beta \neq \alpha$ is assumed.

\begin{table}
\begin{tabular}{c || c | c | c | c }
    \#   &  $J_1$ &  $J_2$    &  $J_3$   &  $J_4$    \\\hline
    I    &  $1$   & $0$       &    $0$   &  $-0.24$  \\
    II   &  $0.5$ & $0.25$    & $-0.125$ &  $-0.24$  \\
    III  &  $2$   & $-0.5$    & $0.25$   &  $-0.24$  \\
\end{tabular}
\caption{Three different sets of lattice exchange parameters (in a.u. "$J_0$") corresponding to the same micromagnetic model. 
To calculate the lifetime of hopfions, we take $J_0$=10 meV.}
\label{tab1}
\end{table}

\begin{figure*}
    \centering
    \includegraphics[width=0.95\textwidth]{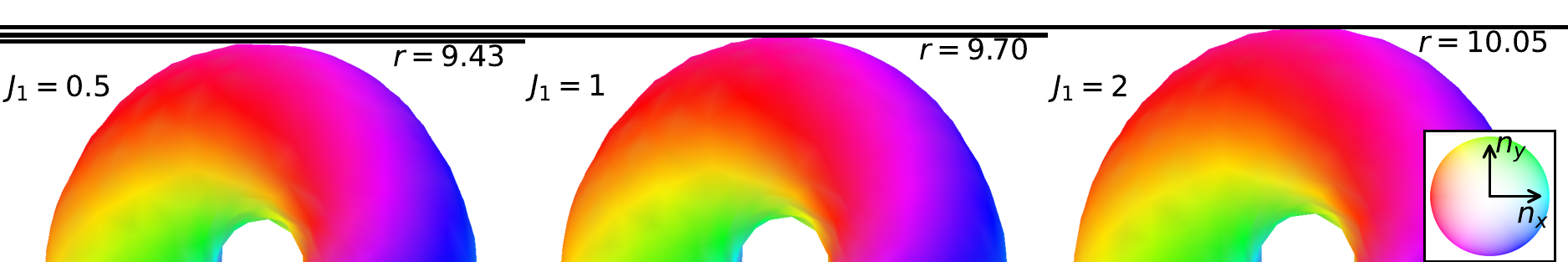}
    \caption{Hopfions for 3 sets of exchange integrals given in Table \ref{tab1}, which corresponds to the same micromagnetic parameters.
    The equilibrium hopfion radii shown in the figure are slightly different.
    Values $J_1$ are given in units $J_0$ and radii $r$ in units of lattice constant $a$.}
    \label{fig:radii}
\end{figure*}

The correspondence between the parameters of the micromagnetic and discrete models is ambiguous. 
One and the same set $\mathcal{A}, \mathcal{B}, \mathcal{C}$ may correspond to several sets of exchange integrals $J_n$.
For a simple cubic lattice, taking into account the interaction up to the fourth nearest neighbors ($n=4$)
the micromagnetic  parameters and exchange constants $J_1, J_2, J_3, J_4$ are connected by linear relations:
\[
a\mathcal{A} = \frac{J_1}2 + 2(J_2 + J_3 + J_4),
\]
\[
\mathcal{B}=-a\left(\frac{J_1}{96}+\frac{J_2+J_3}{24}+\frac{J_4}{6}\right),
\]
\[
\mathcal{C}=-a\left(\frac{J_1}{48}+\frac{J_2}{3}+\frac{7 J_3}{12}+\frac{J_4}{3}\right),
\]
where $a$ is the lattice constant,
Below we consider three sets of lattice exchange integrals, displayed in Table \ref{tab1},
in arbitrary units ($J_0$) corresponding to the same micromagnetic parameters:
$a\mathcal{A}=0.02$ ($J_0$), $\mathcal{B}/a=0.02958(3)$ ($J_0$), $\mathcal{C}/a=2\mathcal{B}$ ($J_0$).
It is convenient to introduce two parameters that affect the shape and size of hopfions, 
which will be studied below:
\[
r_0=\sqrt\frac{\mathcal{B}+\mathcal{C}}{\mathcal{A}}\approx 2.1,\quad 
\gamma_0=\frac{\mathcal{B}}{\mathcal{B}+\mathcal{C}}=\frac23.
\]
The first parameter, which has the dimension of length, isotropically changes the scale of localized structures, the second is responsible for their anisotropy. An increase in $\gamma_0$ also reduces the characteristic size of the structure \cite{Sallermann22}.

The hopfion stability criterion \cite{Rybakov22}
\[
max(\mathcal{C},6\mathcal{B})\geq 6.5\mathcal{A} a^2,
\]
is satisfied for the parameters presented in the Table \ref{tab1},
however, we close to the boundary of the stability region, 
hence we work with fairly  small hopfions and lattice effects may be important.

Hopfion simulation in a discrete lattice is performed using the standard Heisenberg model. 
Denoting by $\bm n_i$ the unit vector along the magnetic moment on the $i$-th site, the energy of the system can be written as
\[
E = -\sum_{\langle i,j\rangle} J_{ij} \bm n_i\cdot\bm n_j,
\]
where the summation is over all pairs $\langle i,j\rangle$ of magnetic moments in simple cubic lattice. 
The exchange constants $J_{ij}$ coincide with $J_s$ introduced above if $j$ belongs to the $s$-shell of $i$, and it is assumed that $J_s=0$ for $s>4$.
The simulation domain consists of $60\times60\times60$ moments, which is about twice the size of hopfion under consideration. 
To reduce the influence of the boundary of the simulation domain,
we pin all the moments on the faces of the cube, assuming that they are oriented along the $\hat z$ axis.

The metastable hopfion state is obtained by minimizing the energy from 
the corresponding ansatz (see Appendix C.III in \cite{Rybakov22}) 
using the nonlinear conjugate gradient method \cite{Fischbacher17}, \cite{Lobanov-CPC21}. 
For all the parameters specified in Table \ref{tab1}, the hopfion in the ferromagnetic (FM) phase is metastable and its symmetry axis is directed along the (1, 1, 1) crystallographic axis due to lattice effects.

The hopfion simulation in the bulk of the sample was carried out using periodic boundary conditions both with and without spin pinning on the surface of the simulated cell. 
Without pinning, the hopfion has  quasi-zero modes, which correspond to the quasi-zero eigenvalues of the  hessian of energy. 
The pinning at the boundaries removes all zero modes.
Except for this, the simulation results with and without pinning are identical.

Fig. \ref{fig:radii} shows the surface $n_z$=0 for hopfions obtained with the values of the parameters given in Table~1, which correspond to the same micromagnetic system. 
Although the skyrmions are quite similar, their sizes are slightly different, increasing with the value of $J_1$. 
The axis of symmetry for all three hopfions corresponds to the most energetically favorable direction $(1,1,1)$.

\section{ Minimal energy path for hopfion collapse and escape trough the boundary}

\begin{figure*}
    \centering
    \includegraphics[width=0.95\textwidth]{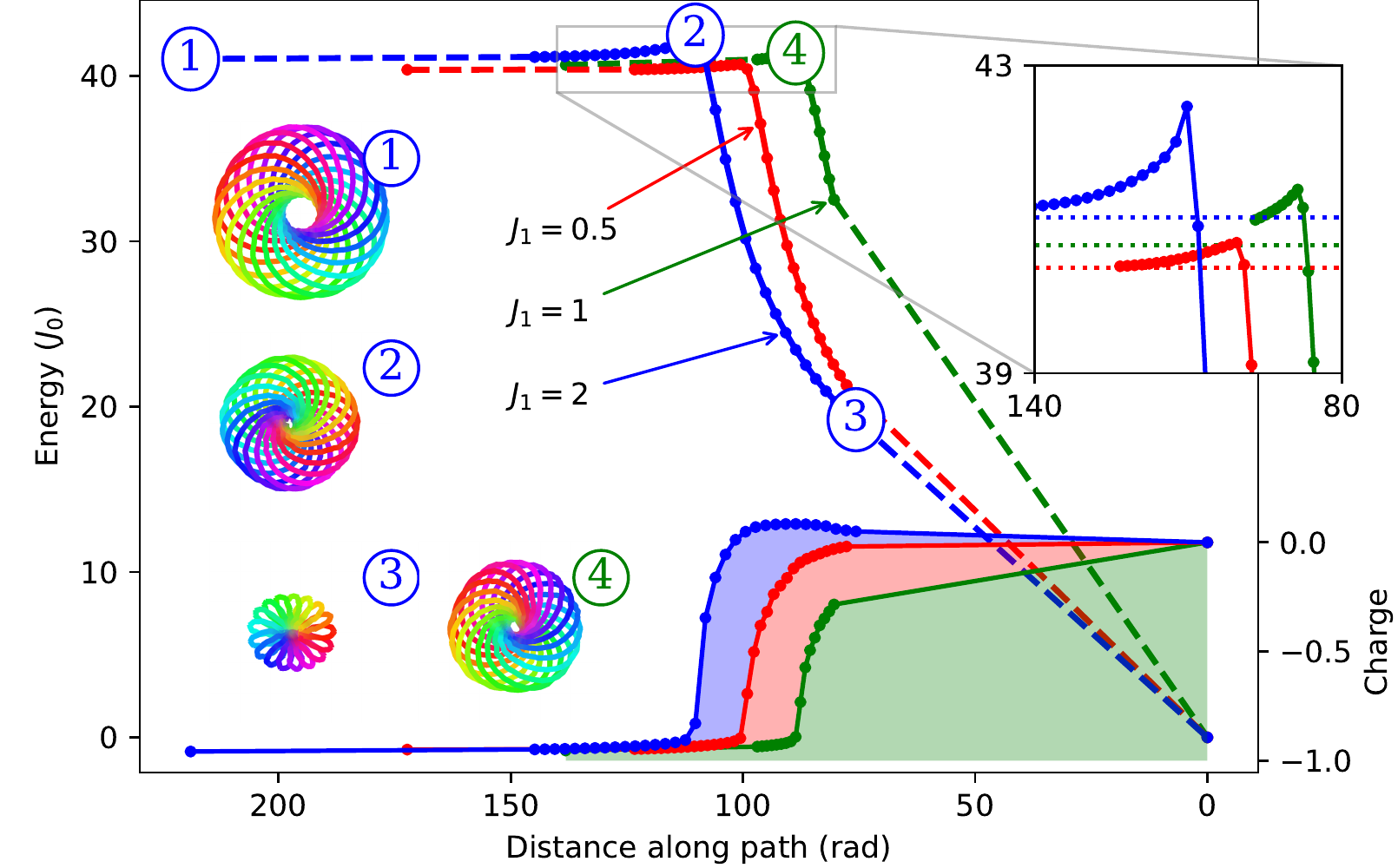}
    \caption{MEP for the decay of hopfions inside the sample and magnetic configurations at selected points.
     The insets show the paths in the vicinity of the saddle points.
     On the lower graphs and on the right axis, the hopfion charge is along the trajectory.
     Magnetic texture states along MEP are illustrated by lines of constant orientation of the
     magnetic moments with vanishing z-projection of the moments using the same color scheme as in Fig. \ref{fig:radii}. 
     }
    \label{fig:mep_decay}
\end{figure*}

\begin{figure*}
    \centering
    \includegraphics[width=0.95\textwidth]{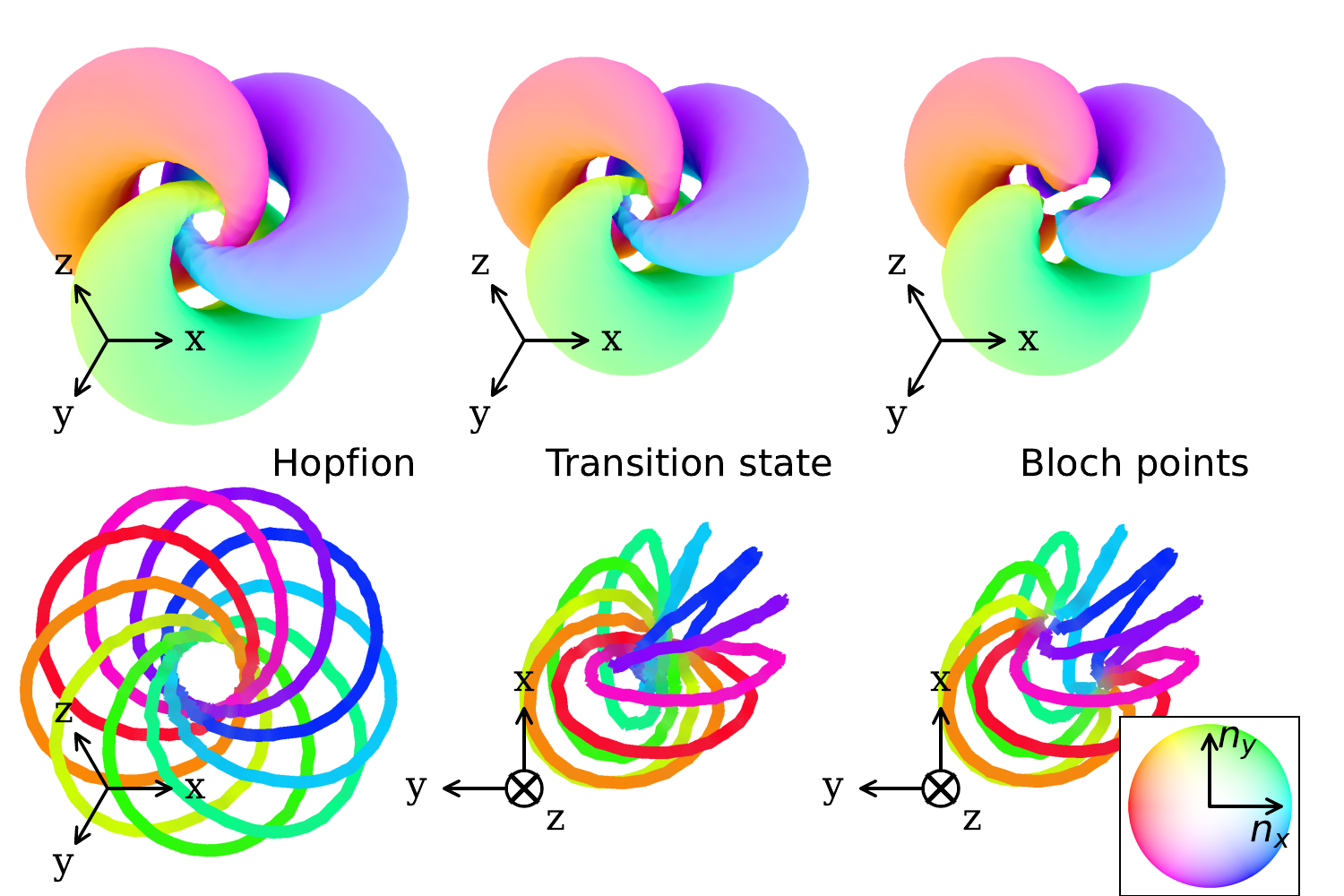}
    \caption{Hopfion (left), transition state (middle) and a state halfway after transition state to FM state
    having two Bloch points (right) for $J_1=0.5$.
    Top row demonstrates surface of constant angle $\pi/5$ between magnetic moments and three vectors $(\cos\alpha,\sin\alpha,0)$,
    $\alpha=0,\pm2\pi/3$.
    Bottom row shows lines of constant orientation of the magnetic moments with vanishing $z$-projection of the moments.
    }
    \label{fig:states}
\end{figure*}

\begin{figure*}
    \centering
    \includegraphics[width=0.95\textwidth]{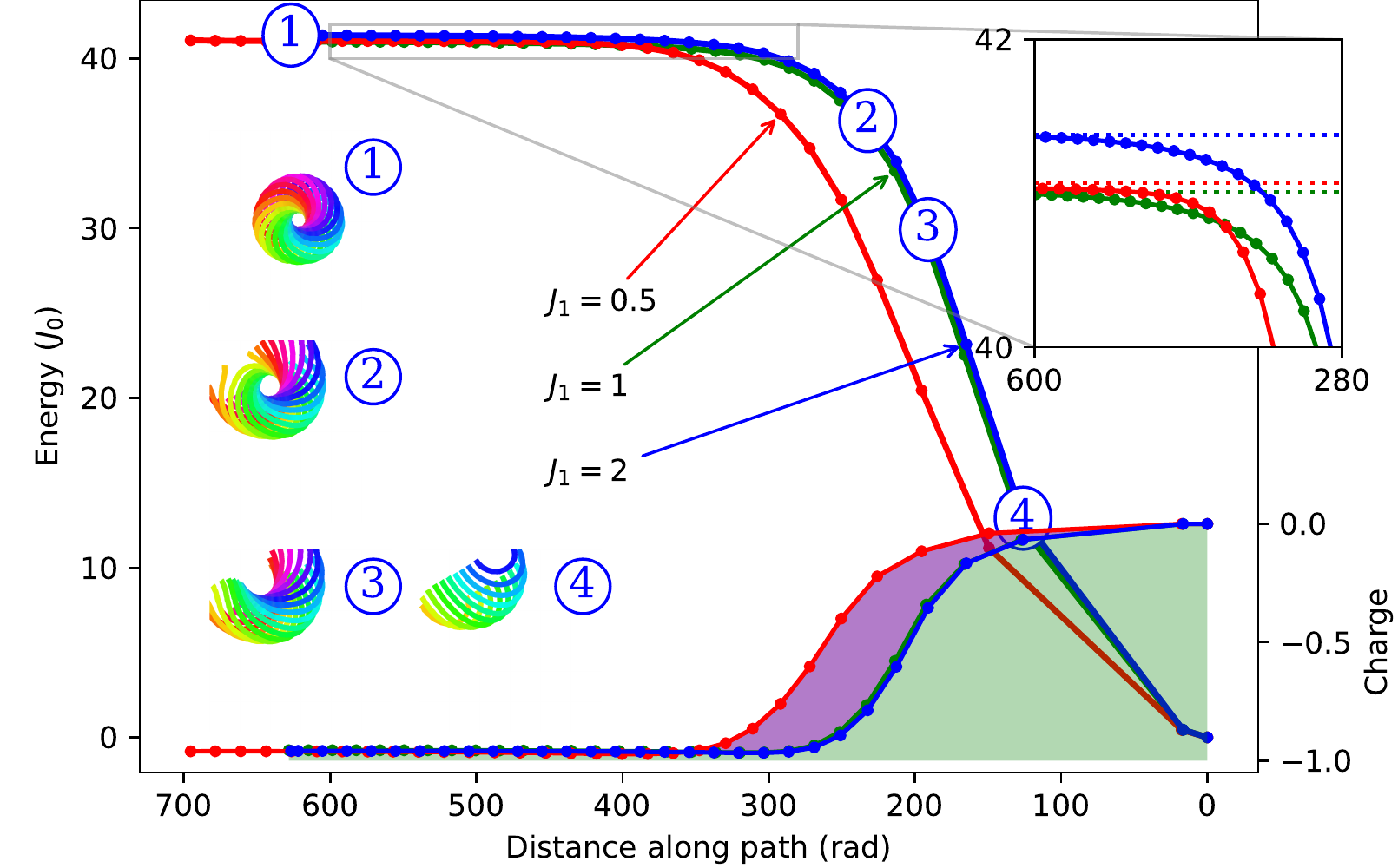}
    \caption{The MEP for hopfion  escape through the boundary and magnetic configurations at selected points.
    There is no energy barrier for this process. 
     The insets show the paths in the vicinity of the saddle points.
     On the lower graphs and on the right axis, the hopfion charge is deposited along the trajectory.
     Magnetic texture states along MEP are illustrated by lines of constant orientation of the
     magnetic moments with vanishing z-projection of the moments using the same color scheme as in Fig. \ref{fig:radii}. 
     }
    \label{fig:mep_escape}
\end{figure*}

The lifetime of magnetic states can serve as a quantitative measure of their stability with respect to thermal 
fluctuations and random external perturbations. These lifetimes or rates of magnetic transitions can be estimated 
using the transition state theory for magnetic degrees of freedom \cite{Bessarab12,Lobanov21}. 

Within the framework of this approach the energy surface is considered as a functional of variables that completely determine the magnetic configuration. 
Local minima on this surface correspond to the ground (FM) and hopfion states. 
Knowing these states, one can find the minimum energy path (MEP) between them. 
The maximum energy along the path is reached at a saddle point on the energy surface. 
Then the difference between the energies of the saddle point $E_{sp}$ and the initial equilibrium hopfion state $E_h$ gives the activation energy of the hopfion collapse $\Delta E_c=E_{sp}-E_h$. 
The activation energy of hopfion nucleation is the difference between $E_{sp}$ and the energy of the FM state $E_f$: $\Delta E_n=E_{sp}-E_f$.

There are various methods for finding MEPs \cite{Lobanov21,Ivanov20}.
By definition, MEP is a path in phase space, such that it starts and ends at energy minima, and each path point is a local energy minimum in a subspace orthogonal to the path at that point.
The path is represented by a set of discrete replicas of the system $\bm n^{(k)}$, called  images, that provide a discrete representation of the path that initially starts with some interpolation between initial and final states and then converges to MEP using some iterative optimization method.
Each iteration of these methods starts by calculating the energy anti-gradient on the path images:
\[
\bm g^{(k)} = -\frac{\partial E[\bm n^{(k)}]}{\partial \bm n},
\]
and the projection of the gradient onto the tangent space to
manifold associated with constraints on the value of magnetic moments:
\[
{\bm q^{(k)}_i} = \bm g^{(k)}_i - \bm n^{(k)}_i(\bm g^{(k)}_i\cdot \bm n^{(k)}_i). 
\]
The vector $\bm q^{(k)}$ is equal to zero for all stationary points,
and determines the direction of the fastest decrease in energy at each image.
Then $\bm q^{(k)}$ projects into the space orthogonal to the path:
\[
\bm p^{(k)} = \bm q^{(k)} - \bm t^{(k)} (\bm t^{(k)}\cdot \bm q^{(k)} ),
\]
where $\bm t^{(k)}$ is the tangent to the path on the image $\bm n^{(k)}$.
The tangent $\bm t^{(k)}$ is estimated by weighted finite difference of the images,
but it is essential to use stable estimate to obtain convergence \cite{Henkelman2000}.
On a MEP, the vector $\bm p^{(k)}$ should be equal to zero, and moreover a small variation in the position of the image in the direction 
orthogonal to $\bm t^{(k)}$ should locally increase the energy of the images.
However, to obtain a good estimation of the transition state the path should be well resolved in a vicinity of the maxima.
The condition is commonly enforced ensuring constant distance between images and use of the climbing image method \cite{Henkelman2000-CI}.
The distance between images can be controlled introducing an elastic force as done in nudged elastic band (NEB) method \cite{Bessarab2015}.
We adopted another common approach called string method, which does not introduce arbitrary parameters. 
In the string method the force $\bm p^{(k)}$ is not modified and auxiliary images are updated as in gradient descend method:
\[
 \bm {\tilde n}^{(k)} = \bm n^{(k)} + \eta \bm p^{(k)},
\]
where step size $\eta$ can in most cases be a constant.
After that, the continuous path is approximated by a spline that goes via the images $\bm {\tilde n}^{(k)}$.
Our experience shows that a piecewise linear approximation is enough.
Then images defining new approximation of MEP are updated in such a way that 
all $\bm n^{(k)}$ belongs to the spline and are equidistant.
The climbing image algorithm can be combined with a string method in a similar way as NEB method \cite{Lobanov21}. We will use this approach below.

To determine the lifetimes of magnetic states within harmonic  transition state theory, it is important to know the topography of the energy surface in the vicinity of the saddle point. In this case,  it is not necessary to determine the entire MEP. This makes it possible to use the truncated MEP method \cite{Lobanov2017}, which allows finding only the part of the path that includes the saddle point. This method has been used to search for the MEP for hopfion collapse. Calculations were performed using string method in Cartesian coordinates
\cite{Lobanov-CPC21} starting optimization from an ansatz.

The ansatz for hopfion annihilation was the geodesic path between the hopfion and the ferromagnetic state.
For Hopfion escape through the boundary, the initial path consists of copies of the Hopfion states translated by several lattice constants.
The part of the hopfion that goes beyond the boundary is discarded,
spins coming out of the boundary are set parallel to the FM state.

Fig. \ref{fig:mep_decay} shows the MEP for three hopfion states corresponding to the same state in the continuous micromagnetic model. The initial magnetic configurations represented in Figs. \ref{fig:radii} have slightly different energy. The larger the hopfion, the greater its energy in the equilibrium state. The axis of symmetry for all states is (111). 
In the figure, the energy along the path is plotted as a function of the reaction coordinate, which is chosen as total rotation angle of all magnetic moments along the path. Since the final FM state is the same for all structures, the reaction coordinate is measured from this state. The distances to the equilibrium hopfion state along the MEP are different for the three hopfion structures. The calculations shown in the figure by solid lines were carried out for the part of the path, including the saddle point, using the truncated MEP method. The dashed lines correspond to transitions to the initial and final states from the ends of the truncated MEP.

The MEP of hopfion collapse consists of several stages. 
First, the hopfion shrinks to the state where its inner radius is about a few lattice constants. 
This state corresponds to the saddle point on the energy surface. 
In this case, for the hopfion III ($J_1=2$), the symmetry axis at the saddle point changes to (110), 
while for I and II it remains directed along (111). 
We estimated hopfion charge for all computed images by a quadrature approximation of Whitehead formula \cite{Whitehead47}.  
The hopfion charge begins to decrease as an image approaches the saddle point, 
as shown in the lower part of the fig \ref{fig:mep_decay}. 
At the next stage, in the vicinity of the transition state, 
two Bloch points are formed in the center of the hopfion, 
and move in opposite directions, leading to the formation of a toron-like structure. 
Radius of the toron is close to the outer radius of the hopfion.
At the final stage, the toron decays by shrinking, with both its radius and the distance between the Bloch points tending to zero.

Fig. \ref{fig:states} shows the hopfion equilibrium state, the structure in the vicinity of the saddle point on the energy surface and outside the transition region, where the Bloch points have formed and have already diverged. It is this sequence of states that occurs when moving along the MEP.

Another mechanism for the removal of hopfion from a sample of finite size is the hopfion escape through the free boundary. 
Analogous processes with skyrmions require an activation energy, since the skyrmion is repelled from the free boundary \cite{Uzdin18}. 
The MEPs for hopfion crossing the boundary are shown in fig. \ref{fig:mep_escape}.
None of the considered hopfions has an energy barrier for crossing the border. 
The hopfion charge monotonically decreases in absolute value as more and more of the non-collinear magnetic structure leaves the sample.
To confine the hopfion inside the sample boundary conditions should be modified,
e.g. orientation of the magnetic moments can be set constant on the surfaces
as for homeotropic boundary conditions common for liquid crystal films.
An easy axis magnetic anisotropy on the boundary, that can be considered as a weak form of such boundary conditions,
was used for soliton stabilization in \cite{Khodzhaev22,Wang19,Li22}.
Other approaches to stabilization of hopfion are possible,
for example, by embedding in a helical or conical background 
of chiral magnets \cite{Voinescu20} instead of FM phase.

\begin{figure*}
    \centering
    \includegraphics[width=0.95\textwidth]{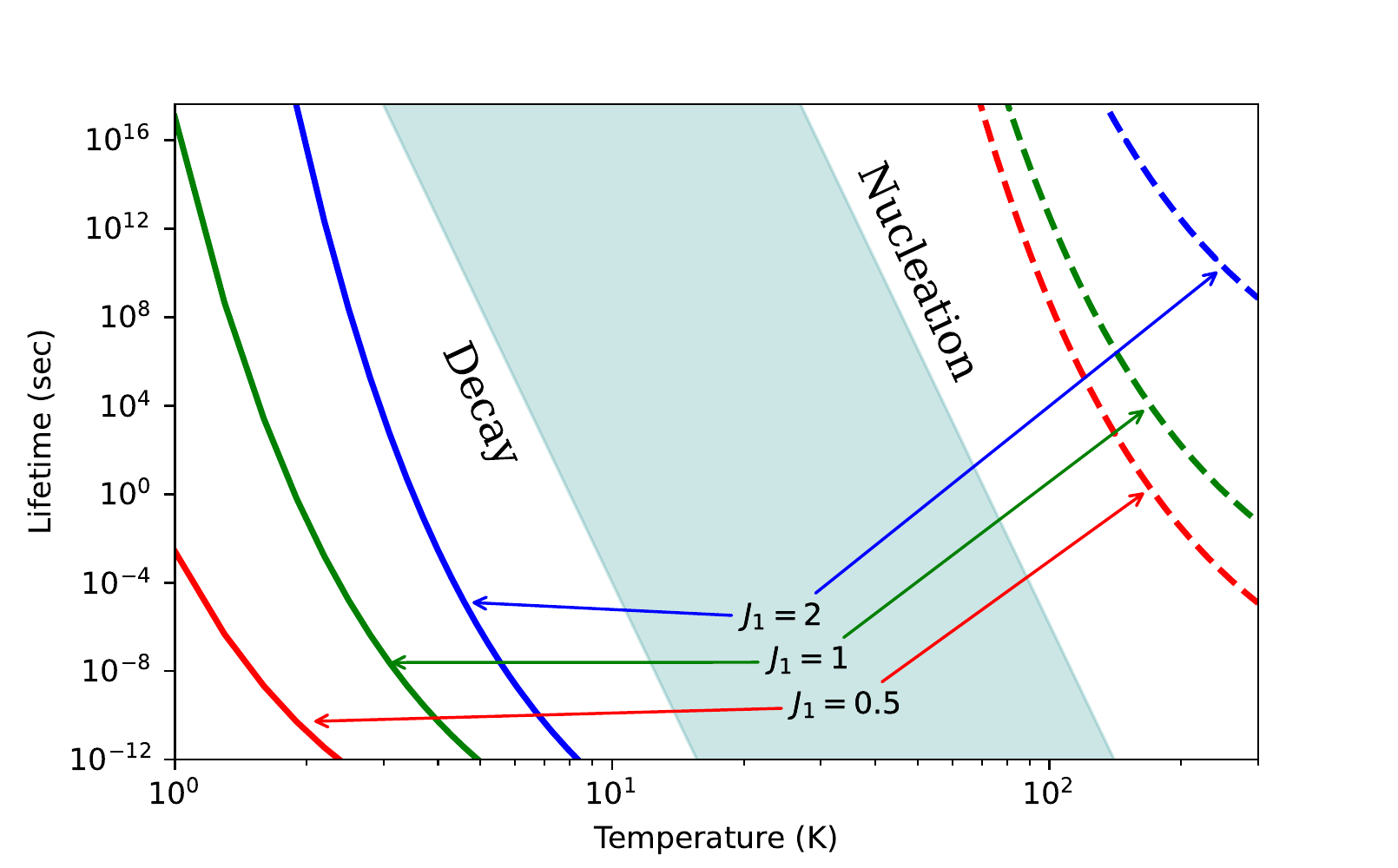}
    \caption{
    Hopfion lifetimes (solid lines) and hopfion nucleation times from the ferromagnetic state (dashed lines) 
    for three sets of parameters from Table \ref{tab1}.
    Hopfion decay is many orders of magnitude more probable than its nucleation for all the parameters.
    }
    \label{fig:lifetime}
\end{figure*}

\section{Pre-exponential factor and lifetime of hopfion states}

\begin{table}
\begin{tabular}{c || c | c | c | c  }
    \#   &  $\Delta E_c$ (meV) &  $\tau_{dyn}$ (ps)   &  $\tau_{ent}$   &  $\tau_0$  (sec)   \\\hline
    I    &  $7.232$    & $1.292$     & $3.148\cdot 10^{-08}$   &  $4.067\cdot 10^{-20}$  \\
    II   &  $3.234$   & $3.787$    & $3.544\cdot10^{-08}$      &  $1.342\cdot 10^{-19}$  \\
    III  &  $14.41$   & $0.71$    & $2.795\cdot 10^{-09}$      &  $1.984\cdot 10^{-21}$  \\
\end{tabular}
\caption{ Hopfion collapse activation energy ($\Delta E_c$), dynamic ($\tau_{dyn}$), entropy ($\tau_{dyn}$) and total pre-exponential factors for three different sets of lattice exchange parameters, listed in the Table \ref{tab1}. $J_0$ is taken equal to 10 meV}
\label{tab2}
\end{table}

Within the framework of harmonic approximation for the shape of the energy surface near the minima and the saddle point, one can obtain the Arrhenius law for the hopfion lifetime \cite{Bessarab12,Lobanov-CPC21}.
\[
\tau = \tau_0 \exp\left(\frac{\Delta E_c}{k_B T}\right),
\]
The activation energy $\Delta E_c$ is determined from the MEP. The pre-exponential factor can be written as the product of the dynamic and entropy parts:
\[
\tau_0 = {2\pi} \tau_{dyn}\tau_{ent}.
\]
The dynamic prefactor $\tau_{dyn}$ depends only on dynamics in the vicinity of the transition state:
\[
\tau_{dyn} = \frac{\mu}{\gamma}\frac{1}{\sqrt{\bm b\cdot\mathcal{H}^{ts}\bm b}},\quad 
\bm b_i = \bm n_i^{ts}\times \bm e_i,
\]
where $\bm n_i^{ts}$  is the spin configuration at the saddle point, $\mathcal{H}^{ts}$ is the Hessian of energy 
in this point, and $\bm e_i$ is the unit eigenvector corresponding to the only negative eigenvalue of the 
operator $\mathcal{H}^{ts}$. $\gamma$ and $\mu$ are the gyromagnetic ratio and the magnetic moment per site, respectively.

The entropy prefactor is the square root of the ratio of the modulus of the Hessian determinant at the saddle 
point $\mathcal{H}^{ts}$ and determinant in the minimum corresponding to the equilibrium hopfion $\mathcal{H}^{min}$
\[
\tau_{ent} = \sqrt{\frac{\det \mathcal{H}^{ts}}{|\det \mathcal{H}^{min}|}}.
\]
Calculating the entropy factor for systems containing hundreds of thousands of atoms is a complex computational problem. For the system under consideration, this problem can be solved due to the short-range exchange interaction using LU decomposition for block band matrices
\cite{Lobanov-CPC21}.

Table \ref{tab2} presents the results of calculations of the dynamic, entropy, and total pre-exponential factors of hopfion collapse, 
as well as the activation energy of this process. 
It differs significantly for the considered sets of parameters corresponding 
to the same micromagnetic models. The lifetime of hopfions as a function of temperature is shown in Fig \ref{fig:lifetime}.
In the calculations $ J_0 =10\ meV$ and $\mu=3\ \mu_B $.

Only the hopfion corresponding to J=2 is stable at T = 2K. This is due to its relatively high activation barrier. The same figure shows the time of generation of hopfions due to temperature fluctuations as a function of temperature.
Such processes can occur at temperatures of hundreds of K, when the equilibrium hopfions are unstable. Therefore, hopfions can only be created artificially, by means of a special external influence, and observed at ultralow temperatures.

The computed prefactor $\tau_0$ for the Hopfion lifetime represented in Table \ref{tab2} is 
on the order of $10^{-20}$s.
The inverse, often referred to as ``attempt frequency'' for the collapse rate, is about $10^{20}s^{-1}$ and 
differs by $8-10$ orders of magnitude from the typical pre-factor for other topological structures, 
such as 2D magnetic skyrmions. 
It is worth noting that in the general case, the attempt frequency is the frequency of the physical process, 
that sets the time scale, but it is a quantity depending on the shape of the energy surface 
in the vicinity of the saddle point and initial state.
For example, in the case of magnetization reversal of small Fe islands the  W(110) surface, 
the attempt frequency within the TST turned out to be $10^{14}-10^{18}$ \cite{Bessarab13} 
and agrees well with scanning tunneling microscopy measurements \cite{Krause09}.
Therefore, in nanomagnetism, such an order of magnitude of $\tau_0$ is not very surprising.

In our calculations, we use the harmonic approximation for the shape of the energy surface. 
If the system has zero modes in the initial and transition states, 
then the harmonic approximation can be violated along the corresponding directions on the energy surface. 
However, in our calculations zero modes were removed by pinning of boundary magnetic moments.
Therefore, the result for $\tau_0$ is not related to limitations of harmonic TST, 
but reflects the actual behavior of the hopfions.

The small value of the prefactor for the hopfion lifetime is related to the large entropy of 
the structure in the transition states compared to the hopfion entropy in the initial state. 
For a 2D skyrmion, on the contrary, the skyrmion has a greater entropy 
than the transition state structure. 
This explains the difference in the pre-exponential factors.

\section{Conclusions}
Hopfion states are local energy minima in three-dimensional magnetic systems with competing exchange interactions. In a continuous micromagnetic model, the orientation of the hopfion symmetry axis does not affect its properties. On a discrete lattice, such a dependence arises due to the appearance of certain, specific directions of the crystallographic axes. One and the same continuous model can correspond to several discrete systems with different sets of exchange parameters. Due to lattice effects, their stability can vary greatly.

Although the equilibrium shape and size of hopfions are quite close for all three sets of parameters, the activation energy and the path leading to hopfion collapse are different. The pre-exponential factors in the Arrhenius law obtained in the harmonic approximation of the transition state theory also differ greatly. Calculations show that hopfions of such a scale can exist as excitations above the FM phase only at ultralow temperatures of a few degrees K. With an increase in the hopfion size, lattice effects should disappear, and one can expect that the hopfion will be more stable and independent of the specific parameters of the discrete model, corresponding to specific micromagnetic parameters.
Increase of $J_0$ or hopfion size can sufficiently increase stability.

The energy barriers for hopfion collapse corresponding to different values of the micromagnetic parameters 
$\mathcal{A}, \mathcal{B}, \mathcal{C}$ were calculated by Sallerman et al. \cite{Sallermann22} based on the 
analysis of the energy surface. They showed that the energy barrier can reach a considerable height and is 
largely determined by the hopfion size with respect to the lattice constant.
In our calculations, the chosen parameters correspond to only one set $\mathcal{A}, \mathcal{B}, \mathcal{C}$, 
whose values are close to the stability limit of the hopfion states, and the size of the hopfion turns out to be quite small.
The pre-exponential factor in the Arrhenius law was not calculated in \cite{Sallermann22}, 
but calculations we have carried out for 2D skrymions have shown
that the pre-exponential factor can increase by $4-5$ orders of magnitude with an increase 
in the radius of the skyrmion by $40$ times relative to the lattice constant \cite{Potkina22}. 
Therefore, it is reasonable to expect that hopfions of larger radius can be stable at higher temperatures, 
which agrees with \cite{Sallermann22}.

\vspace{1cm}

\section{Acknowledgments}
The authors thank N. S.~Kiselev, F. N.~Rybakov, S.~Bl\"ugel and H. J\'onsson for fruitful discussion.
The study was supported by the Russian Science Foundation 
grant No. 22-22-00565, \\ https://rscf.ru/project/22-22-00565/

\bibliography{bibliography-3}

\end{document}